\documentclass[twocolumn, superscriptaddress]{revtex4}

\usepackage{graphicx}
\usepackage{bm}

\newcommand{\beq}{\begin{equation}}
\newcommand{\eeq}{\end{equation}}
\newcommand{\beqa}{\begin{eqnarray}}
\newcommand{\eeqa}{\end{eqnarray}}

\newcommand{\ket}[1]{\left| #1 \right\rangle}

\newcommand{\natc}{N@C$_{60}$}

\begin{document}

\title{\emph{Bang-bang} control of fullerene qubits using ultra-fast phase gates}

\author{John~J.~L.~Morton}
\email{john.morton@materials.ox.ac.uk} \affiliation{Department of
Materials, Oxford University, Oxford OX1 3PH, United Kingdom}
\affiliation{Clarendon Laboratory,
Department of Physics, Oxford University, Oxford OX1 3PU, United
Kingdom}

\author{Alexei~M.~Tyryshkin}
\affiliation{Department of Electrical Engineering, Princeton
University, Princeton, NJ 08544, USA}

\author{Arzhang~Ardavan}
\affiliation{Clarendon Laboratory,
Department of Physics, Oxford University, Oxford OX1 3PU, United
Kingdom}

\author{Simon~C.~Benjamin}
\affiliation{Department of Materials, Oxford University, Oxford
OX1 3PH, United Kingdom}

\author{Kyriakos~Porfyrakis}
\affiliation{Department of Materials, Oxford University, Oxford
OX1 3PH, United Kingdom}

\author{S.~A.~Lyon}
\affiliation{Department of Electrical Engineering, Princeton
University, Princeton, NJ 08544, USA}

\author{G.~Andrew~D.~Briggs}
\affiliation{Department of Materials, Oxford University, Oxford
OX1 3PH, United Kingdom}

\date{\today}

\maketitle

\textbf{Quantum mechanics permits an entity, such as an atom, to
exist in a superposition of multiple states simultaneously. Quantum
information processing (QIP) harnesses this profound phenomenon to
manipulate information in radically new ways~\cite{deutsch85}.  A
fundamental challenge in all QIP technologies is the corruption of
superposition in a quantum bit (qubit) through interaction with its
environment. Quantum \emph{bang-bang} control provides a solution by
repeatedly applying `kicks' to a qubit~\cite{gadiyar94,viola98},
thus disrupting an environmental interaction. However, the speed and
precision required for the kick operations has presented an obstacle
to experimental realization. Here we demonstrate a phase gate of
unprecedented speed~\cite{vandersypen,berryqc1} on a nuclear spin
qubit in a fullerene molecule, and use it to bang-bang decouple the
qubit from a strong environmental interaction. We can thus trap the
qubit in closed cycles on the Bloch sphere, or lock it in a given
state for an arbitrary period. Our procedure uses operations on a
second qubit, an electron spin, in order to generate an arbitrary
phase on the nuclear qubit. We anticipate the approach will be vital
for QIP technologies, especially at the molecular scale where other
strategies, such as electrode switching, are unfeasible.}

Two well known concepts in overcoming the corruption of information
stored within a qubit are decoherence free subspaces~\cite{duan97,
zanardi97, lidar98, kielpinski01}, and quantum error correcting
codes~\cite{shor95,steane96,knill2005}. The former is the passive
solution of restricting oneself to some set of states that, due to
symmetries in the system, are largely immune to the dominant types
of unwanted coupling. The latter is a sophisticated form of feedback
control whereby the effect of unwanted coupling is detected and
corrected. Between these limits there is the idea of dynamical
suppression of coupling --- making some rapid, low level
manipulation of the system so as to actively interfere with the
decoherence process.  Ideas here often relate to the `quantum Zeno
effect' in which repeated measurement (or some related process) is
capable of suppressing the natural evolution of the
system~\cite{italianqze,itano90}. As the system evolves from one
quantum eigenstate, $\ket{0}$, to another, $\ket{1}$, it passes
through a superposition state $\alpha\ket{0}+\beta\ket{1}$ which,
when measured, collapses to one of the two eigenstates with
probabilities given by how far the system has been allowed to
evolve. Therefore, if the measurements are made often, the system
will have a high probability of being locked in the starting state
(though, eventually, the finite probability of flipping will be
realised).

\begin{figure}
\includegraphics[width=8.8cm]{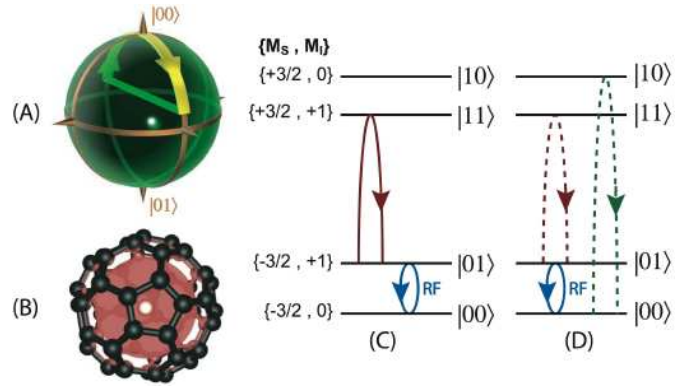}
\caption{A representation of our decoupling scheme, and the physical
system to which it is applied. (a) Our decoupling process --- shown
by the path of the nuclear qubit on its Bloch sphere. The line
within the Bloch sphere is a visual guide and does not indicate that
the state becomes mixed. The state leaves the two-state space
represented by the Bloch sphere, and returns at the indicated point
on the opposite side, remaining pure at all times. An RF field is
applied to drive state $\ket{00}$ to $\ket{01}$. However, applying a
phase of $-1$ to $\ket{01}$ during the evolution causes the state to
jump to the other side of the Bloch sphere, from which it must
evolve back to its initial state. The process can be repeated to
prevent the system from ever reaching $\ket{01}$. (b) A model of the
\natc~molecule. (c) Although this \natc~system of S=3/2 electron
spin and I=1 nuclear spin consists of 12 energy levels, we use only
those four illustrated above. Nuclear Rabi oscillations are driven
between $\ket{00}$ and $\ket{01}$, whilst the phase inversion is
provided through a selective microwave pulse between $\ket{01}$ and
$\ket{11}$. (d) By driving both electron spin transitions
simultaneously (with a \emph{single} off-resonance pulse) an
arbitrary phase can be applied between the $\ket{00}$ and $\ket{01}$
states.} \label{elevels}
\end{figure}

In a related technique which is part of the \emph{bang-bang} family
of strategies~\cite{viola98, lidar05}, measurement operations are
replaced by the application of rapid rotations on the superposition.
The original bang-bang literature proposed the use of rapid bit
flips to prevent unwanted phase evolution.  Our implementation is
very close in spirit to that original paper: we perform the logical
complement of the protocol, using rapid phase shifts to prevent
amplitude evolution (see Fig. 1a and supplementary
movie~\footnote{Animated movie available for download at
http://www.nanotech.org/research/nphys}). In many systems, energy
loss is a primary decoherence mechanism; such decoherence would be
suppressed by precisely this strategy.

This technique is distinct from the projective Zeno effect in two
interesting respects. In addition to locking the system in one of
the two eigenstates the same sequence is capable of freezing any
superposition state. Secondly, as the behaviour is always unitary,
it is remains deterministic even when the pulse frequency is
limited. Given ideal pulses, one could trap the state in a closed
cycle indefinitely.

We demonstrate this dynamic decoupling effect using a pair of
coupled nuclear and electron spins in the endohedral fullerene
molecule \natc~(shown in Fig.~\ref{elevels}b)~\cite{Murphy1996}.
Information on the origin of the 12-level spin system in \natc~is
provided in the supplementary online material, along with other
experimental details.  We believe this system is the strongest
candidate for a molecular qubit: the electron spin degree of freedom
permits initialisation to a genuine pure state, while the protection
afforded by the fullerene armour yields the longest decoherence time
measured for any molecular electron
spin~\cite{harneit,briggsRS,eseem03}. For the experiments described
here, we choose four levels from the 12 present in such a way that
they correspond to two qubits: an electron qubit (formed from
$M_S=+3/2, -3/2$) and a nuclear qubit ($M_I=+1,0$), as shown in
Fig.~\ref{elevels}c. As the nuclear qubit has excellent natural
environmental decoupling we must \emph{introduce} a strong coupling
by applying a resonant radio-frequency (RF) field to drive Rabi
oscillations. This field is then successfully decoupled by fast
phase `kicks' to the system: we exploit the qubit-qubit coupling,
taking the electron qubit around closed cycles so as to apply phase
shifts to the nuclear qubit. Finally, we observe that by detuning
the microwave pulse away from the electron spin transition
frequency, this technique can be generalised to apply an arbitrary
phase gate to the nuclear spin on a time scale which is much faster
than normal nuclear magnetic resonance (NMR) methods.

\begin{figure}
\includegraphics[width=9cm]{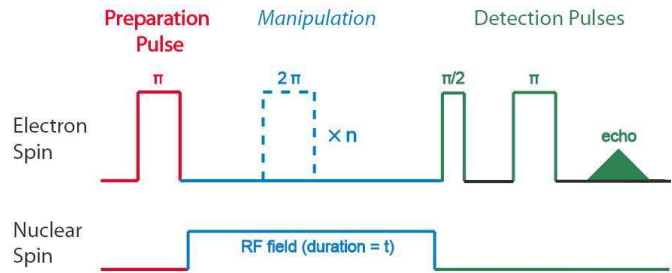}
\caption{The electron-nuclear double resonance (ENDOR) pulse
sequence used.} \label{pulseseq}
\end{figure}

The general pulse sequence used is shown in Fig.~\ref{pulseseq}. A
selective electron $\pi$-pulse on the $M_I$=+1 manifold transfers
the thermal polarisation of the electron spin to the nuclear spin
(this can be thought of as a controlled-NOT operation with the
nucleus as control and electron as target~\cite{mehring03}). The
nuclear polarisation then exhibits Rabi oscillations between the
$M_I$=0 and $M_I$=+1 levels (in Fig.~\ref{elevels} $\ket{00}$ and
$\ket{01}$, respectively) upon the application of a suitable RF
driving field. During the Rabi oscillations, fast microwave pulses
are applied on the electron spin to suppress the nuclear spin
evolution. Finally, a measurement of the relevant state populations
is performed using two-pulse electron spin echo detection.

It is possible to apply a phase gate to a qubit by rotating one of the basis states around a complete cycle through an auxiliary level (Fig.~1c).
This can be simply seen by evaluating the
operator for a complete on-resonance rotation applied to a S=1/2
particle:
\[U_{2\pi}= e^{-{\mathrm i}\frac{1}{2}\sigma_x 2\pi} = \left(
\begin{array}{cc}
  -1 & 0 \\
  0 & -1 \\
 \end{array}
\right),
\]
where $\sigma_x$ is the Pauli spin matrix. Therefore,
when this rotation is selectively performed on the $\ket{01}$ -
$\ket{11}$ transition, state $\ket{01}$ will acquire a $-1$ phase
with respect to $\ket{00}$ (i.e.\ a $\pi$ phase shift). Although the
electron spin used in this case is $S=3/2$, the Bloch vector for
higher spin systems (in high-symmetry environments, such as our
spherical fullerene) behaves identically to the $S=1/2$
case~\cite{schweiger,morton04}.

\begin{figure}
\includegraphics[width=8.8cm]{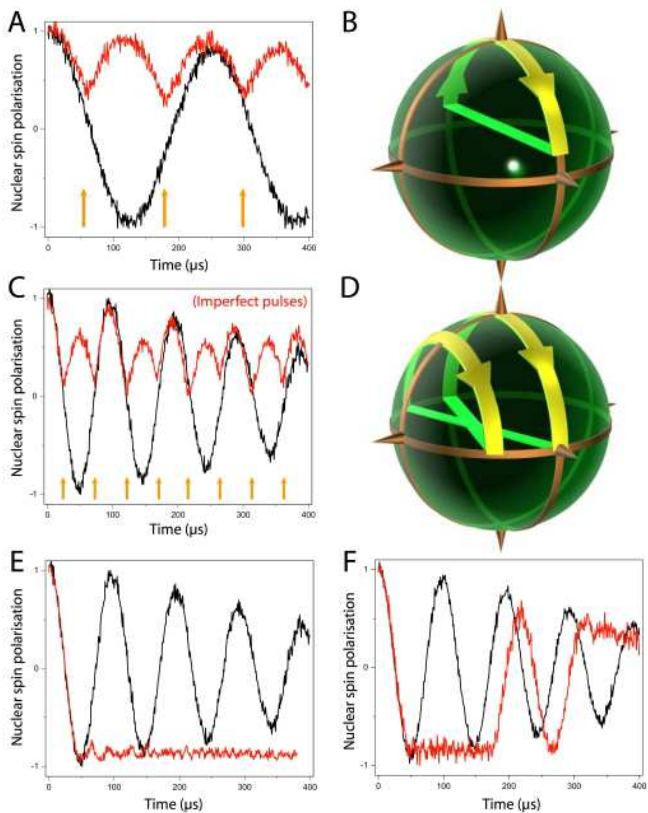}
\caption{The natural evolution between two nuclear spin states of
the nitrogen atom can be disrupted by the application of decoupling
pulses. Unperturbed Rabi Oscillations are shown in black, while
those under the influence of decoupling pulses are shown in red. The
decay observed is an artifact of the inhomogeneity in the RF driving
field, and is not a true decoherence phenomenon. (a,~b)~Microwave
pulses are applied on the electron spin at regular intervals
(indicated by arrows), inverting the phase of one of the nuclear
spin states and reversing the evolution of the nuclear spin qubit.
(c, d) When we implement a phase shift less than $\pi$, an odd-even
behaviour is observed, corresponding to greater and lesser paths on
the Bloch sphere. (e) Increasing the repetition rate of the
microwave pulses locks the system in a particular state. (f) The
qubit can be locked and released at any point.} \label{zenodata}
\end{figure}

Fig.~\ref{zenodata}a shows the effect of applying a $\pi$ phase
shift during the nuclear Rabi oscillations --- the evolution is
reversed each time a microwave pulse is applied, as illustrated on the nuclear Bloch sphere in
Fig.~\ref{zenodata}b. It is not necessary to apply perfect $\pi$
phase shifts in order to lock the spin.  In Fig.~\ref{zenodata}c,
smaller phase shifts are implemented.  After
the first pulse, the nuclear spin evolves around a lesser circle on
the Bloch sphere. A second (identical) phase shift brings the
nuclear spin back onto a great circle, with an overall $\pi$ phase
shift (illustrated in Fig.~\ref{zenodata}d). This kind of error
cancellation is analogous to the
$90^{\circ}_y~180^{\circ}_x~90^{\circ}_y$ type of pulse correction
NMR~\cite{spinchoreo,Levitt1986}.
By increasing the repetition rate of the phase shift pulses, the
nuclear spin evolution can be locked in one particular state
(Fig.~\ref{zenodata}e), and released as desired
(Fig.~\ref{zenodata}f).

\begin{figure}
\includegraphics[width=9cm]{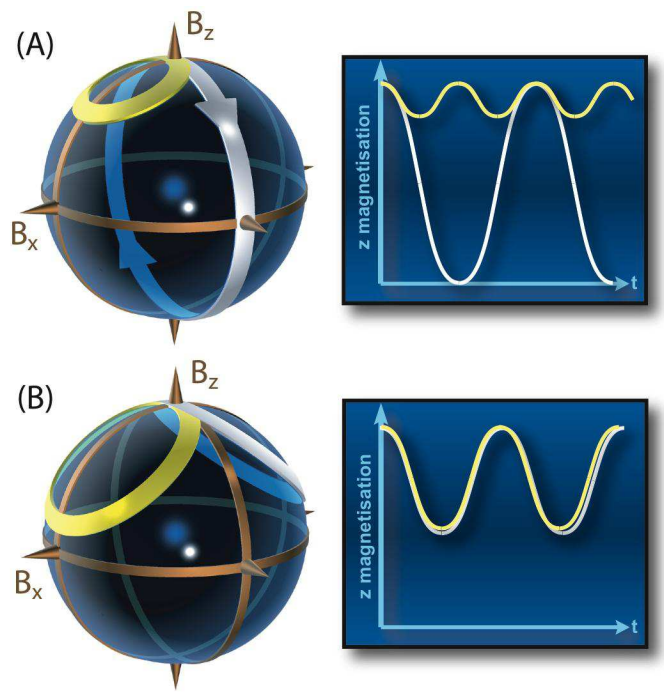}
\caption{Arbitrary nuclear phase gates are implemented by driving two electron spin
transitions simultaneously (see Fig.~\ref{elevels}d). (Left) The path of the electron magnetisation vector driven by a microwave field. (Right) The corresponding z-magnetisation. (a) One transition is driven resonantly, and one strongly detuned. On each cycle, the phase accumulated is proportional to the area enclosed by the path. Both transitions must undergo an integer number of complete cycles to ensure the populations remain unchanged. (b) With
both transitions detuned by equal and opposite amounts, the
population evolution for each is the same, whilst the phase
accumulated is of opposite sign. Hence, the relative phase shift can
be tuned by controlling the microwave pulse power.} \label{offres}
\end{figure}

Any microwave pulse of finite duration which is resonant with the
primary $\ket{01}$-$\ket{11}$ transition also performs a detuned
excitation of the $\ket{00}$-$\ket{10}$ transition (illustrated in
Fig.~\ref{offres}a). The difference between these two transition
frequencies is fixed in our system by the hyperfine coupling constant
$a=15.8$~MHz, but the effective detuning can be controlled through
the microwave pulse power (B1). In the limit of strong selectivity
(weak pulses) only the resonant transition is significantly excited, and a
complete cycle ($\ket{01}$-$\ket{11}$-$\ket{01}$) generates a
$\pi$ phase shift in the nuclear qubit. Faster (non-selective) pulses are possible,
provided that B1 is chosen such that \emph{both} electron transitions
undergo an integer number of complete cycles, thus ensuring the system returns to the subspace $\{\ket{00},\ket{01}\}$ after the
microwave pulse. This permits the implementation of phase shifts which differ from $\pi$, as shown in Figs.~3c and 3d.  The phase acquired by each cycle has a geometric interpretation: it is equal to half of the solid angle subtended by the path of an eigenstate around the Bloch sphere (see Fig.~\ref{offres}a)~\cite{aharonov87}. The differences in the two phases obtained defines the phase shift on the nuclear qubit.

A straightforward extension of this idea is to drive both the
electron spin transitions ($\ket{00}$-$\ket{10}$ and
$\ket{01}$-$\ket{11}$), with equal and opposite detuning (i.e. at a
frequency half-way between their resonances). Here, the condition of
both transitions going through an integer number of complete cycles
can be satisfied for any value of B1. The populations of both
transitions evolve in the same way, whilst the phase acquired is
opposite, as illustrated in Fig.~\ref{offres}b. The relative phase
accumulated (which defines the phase gate applied) is determined
only by B1, allowing an arbitrary phase gate to be applied on the
nuclear spin (see also~\cite{oi05}). The duration of this gate
($\sim$100~ns) is orders of magnitude shorter than typical NMR phase
gates, and about 10$^5$ times shorter than existing geometric phase
gates in NMR~\cite{berryqc1}.

We have demonstrated an ultra-fast fast phase gate on a nuclear spin
qubit, by driving a coupled electron qubit around a closed cycle.
The gate speed can exceed even the nuclear precession frequency,
which is only possible through exploiting the transition selection
rules in the system.  Through repeated application of this fast
phase gate we have bang-bang decoupled a nuclear spin qubit from a
permanent driving field.

The experiments described here were performed on isolated
\natc~molecules in solution, in which the natural noise is low. The
magnitude of interactions in a fullerene-based processor will be at
most that experienced by close-packed \natc~arrays, where the
nuclear spin dipole interaction is of the order 100~Hz, or
approximately 40 times weaker than the RF driving field we apply
here. Paramagnetic impurities within such structures could increase
the natural noise further, however, the speed of the decoupling
gates demonstrated in our experiment show that interactions as
strong as 100~kHz can be suppressed.

This scheme has a broad applicability: the minimum complexity
required of a quantum system is three levels (the two qubit levels
and one auxiliary level) coupled by a suitably rapid allowed
transition. However, as we demonstrate here, the approach also works
with more than three levels and multiple allowed transitions among
those levels. Indeed, a very natural implementation will be any
system with a coupled electron and nuclear spin, giving (in the
$S=1/2$ case) four levels on two different energy scales. For
example, two very different qubits systems also suited to our
approach are phosphorous impurities in silicon, and N-\emph{V}
centers in diamond. Both of these have been identified by the QIP
community as promising qubits. Generally, our demonstration
highlights the potential benefits of physical `qubit' systems beyond
the simple 2-level structure.

In addition to suppressing unwanted coupling to an environment, this
effect is explicitly required in certain quantum computing schemes
(for example, to control the interaction between neighbouring qubits
in perpetually coupled spin-chains~\cite{benjamin05}). Given the
great difficulties associated with tailoring interactions in quantum
systems, it is likely that decoupling strategies of this kind will
form a quintessential element in any real quantum computer.

\section{Acknowledgements}
We acknowledge discussions with Daniel Oi and especially Brendon Lovett. We thank Wolfgang Harneit's group at F.U. Berlin for providing Nitrogen-doped fullerenes, and John Dennis at
QMUL, Martin Austwick and Gavin Morley for the purification of \natc. We thank the Oxford-Princeton Link fund for support. This research is part of the QIP IRC www.qipirc.org. JJLM is supported by St. John's College, Oxford. AA and SCB are supported by the Royal Society. GADB thanks the EPSRC for support (GR/S15808/01). Work at Princeton was supported by the NSF International Office through the Princeton MRSEC Grant No. DMR-0213706 and by the ARO and ARDA under Contract No. DAAD19-02-1-0040. 3D images were created using POV-Ray open source software (povray.org).

\section{Materials and Methods}
The molecular species used in this work is N@C$_{60}$ (also known as
\emph{i}-NC$_{60}$), consisting of an isolated nitrogen atom in the
$^4$S$_{3/2}$ electronic state incarcerated by a C$_{60}$ fullerene
cage.  Our production and subsequent purification of \natc~is
described elsewhere~\cite{mito}. High-purity \natc~powder was
dissolved in CS$_{2}$ to a final concentration of 10$^{15}$/cm$^3$,
freeze-pumped to remove oxygen, and finally sealed in a quartz tube.
Samples were 0.7~cm long, and contained approximately $5\cdot
10^{13}$ \natc~molecules. Pulsed Electron Paramagnetic Resonance
(EPR) measurements were made at 190~K using an X-band Bruker
Elexsys580e spectrometer, equipped with a nitrogen-flow cryostat.

\natc~has electron spin $S=3/2$ coupled to the $^{14}$N nuclear spin
$I=1$. The EPR spectrum consists of three lines centered at electron
g-factor $g=2.003$ and split by a $^{14}$N isotropic hyperfine
interaction $a=0.56$~mT in CS$_2$~\cite{Murphy1996}. The
electron-nucleus double resonance (ENDOR) spectrum consists of four
principle lines associated with the four $M_S$ states, each of which
is further split into two by the second-order hyperfine
interaction~\cite{eseem03}, enabling selective excitation of (for
example) the $M_I$=0 to $M_I$=+1 transition. In these experiments we
applied two simultaneous RF fields (22.598 and 24.782~MHz) to coherently drive both the
$\ket{00}$ - $\ket{01}$ and $\ket{10}$ - $\ket{11}$ transitions.
This is done solely to improve sensitivity, and as all microwave
pulses do not exchange populations between these two subspaces,
they can be treated independently.

The electron spin transition $M_S=+3/2$ to $M_S=-3/2$ is driven via
the intermediate levels $M_S=\pm1/2$, however upon a complete $\pi$
or $2\pi$ rotation, no population remains in the intermediate
levels. This can be straightforwardly seen by evaluating the
rotation operators for high-spin systems.

In the majority of data presented here, microwave pulse power was
chosen so the second off-resonance transition was also excited, such
that  driving the system around one complete cycle generated an
approximate $\pi/2$ phase shift. Thus, a $\pi$ phase shift was
implemented by \emph{twice} driving the system through a complete
cycle.

\end{document}